\documentclass[english,prd,a4paper,onecolumn,nofootinbib]{revtex4}
\usepackage{graphicx, bm, color, babel}
\usepackage{hyperref}
\usepackage{amsmath}
\usepackage{amssymb}
\usepackage{amsfonts}
\usepackage{dsfont}
\usepackage{color}

\def\be{\begin{equation}}
\def\ee{\end{equation}}
\def\bea{\begin{eqnarray}}
\def\eea{\end{eqnarray}}

\newcommand{\<}{\langle}
\renewcommand{\>}{\rangle}

\newcommand\lsim{\mathrel{\rlap{\lower4pt\hbox{\hskip1pt$\sim$}}
        \raise1pt\hbox{$<$}}}

\newcommand{\al}{\alpha}
\newcommand{\ga}{\gamma}
\newcommand{\De}{\Delta}
\newcommand{\de}{\delta}

\newcommand{\ka}{\kappa}

\newcommand{\la}{\lambda}
\newcommand{\dd}{\partial}

\newcommand{\bn}{\mathbf{n}}
\newcommand{\bk}{\mathbf{k}}
\newcommand{\hbk}{\mathbf{\hat k}}
\newcommand{\bx}{\mathbf{ x}}
\newcommand{\by}{\mathbf{y}}
\newcommand{\bl}{\boldsymbol{\ell}}

\newcommand{\avd}[1]{\left\<\,\overline{#1} \,\right\>}

\newcommand{\no}{\mathbf{n}_{o}}
\newcommand{\xo}{\mathbf{x}_{o}}

\newcommand\spart{\;\raise1.0pt\hbox{$/$}\hskip-6pt\partial}
\newcommand\spartb{\;\overline{\raise1.0pt\hbox{$/$}\hskip-6pt
\partial}}

\newcommand{\mR}{{\cal R}}
\newcommand{\mS}{{\cal S}}

\newcommand{\DD}{{\cal D}}

\newcommand{\OO}{{\cal O}}

\newcommand{\bfe}{\mathbf{e}}
\newcommand{\bal}{\boldsymbol{\alpha}}
\newcommand{\bnabla}{\boldsymbol{\nabla}}
\newcommand{\go}{\gamma^{(1)}}
\newcommand{\gt}{\gamma^{(2)}}
\newcommand{\kap}{\langle \kappa_1^2\rangle}
\newcommand{\gl}{\hat\bl^T\!\!\gamma\,\hat\bl}

\begin{document}

\title{\vspace{-0.5cm}{\normalsize \normalfont \flushright CERN-PH-TH-2015-062\\}
\vspace{0.6cm}
Do we care about the distance to the CMB?\\
Clarifying the impact of second-order lensing}
\author{Camille Bonvin$^1$}
\author{Chris Clarkson$^2$}
\author{Ruth Durrer$^3$}
\author{Roy Maartens$^{4,5}$}
\author{Obinna Umeh$^4$}
\affiliation{$^1$Theory Division, CERN, 1211 Geneva, Switzerland\\
$^2$Astrophysics, Cosmology \& Gravity Centre and Department of Mathematcis \& Applied Mathematics, University of Cape Town, Cape Town 7701, South Africa\\
$^3$D\'epartement de Physique Th\'eorique \& Center for Astroparticle Physics, Universit\'e de Gen\`eve, Quai E.\ Ansermet 24, CH-1211 Gen\`eve 4, Switzerland\\
$^4$Physics Department, University of the Western Cape, Cape Town 7535, South Africa\\
$^5$ Institute of Cosmology \& Gravitation, University of Portsmouth, Portsmouth PO1 3FX, United Kingdom\\
{\rm E-mail: camille.bonvin@cern.ch, chris.clarkson@gmail.com, ruth.durrer@unige.ch, roy.maartens@gmail.com, umeobinna@gmail.com}} \vspace*{0.2cm}

\date{\today}

\begin{abstract} 

It has recently been shown that second-order corrections to the background distance-redshift relation can build up significantly at large redshifts, due to an aggregation of gravitational lensing events.  This shifts the expectation value of the distance to the CMB by 1\%. 
In this paper we show that this shift is already properly accounted for in standard CMB analyses. We clarify the role that the area distance to the CMB plays in the presence of second-order lensing corrections. 


\end{abstract}

\maketitle

\section{Introduction}

Second-order perturbations in the standard model are  important for precision cosmology. They carry with them a difficulty in interpretation, because they do not average to zero, and consequently give a (small) change to the background. The most important are potential changes to observables: are these ever significant, and can they lead to biases in cosmological parameter estimation? 
This question is the essence of the problem of backreaction. Analyses in recent years have shown that `backreaction' probably cannot be the origin of dark energy, but it has to be taken into account for precision cosmology as it could bias cosmological parameters at the level of up to several percent, see e.g., \cite{H01,H02,Ben-Dayan:2014swa,Adamek:2015gsa,Clarkson:2011zq,Vanderveld:2007cq,Baumann:2010tm,Wiegand:2011je,Buchert:2011sx}.

Clearly, like all other cosmological quantities the area distance is fluctuating and the power spectrum of its first order perturbations has been determined~\cite{Bonvin:2005ps, hui}. Recently it was shown that {\em second-order} lensing shifts the area distance (also called angular diameter distance) to the CMB. The expectation value $\< d_*\>$ of the area distance to the CMB is larger than the background value $d_{0*}=\chi_*/(1+z_*)$, \cite{cmb}: 
\be\label{ds}
\<d_*\>=d_{0*}\Big[1+\frac{3}{2}\kap\Big],
\ee
where $\kappa_1$ is the first-order magnification at redshift $z_*$. At large distances the shift in the mean area distance builds up proportional to the comoving volume, approximately given by:
\be
\label{kappapp}
\frac{3}{2}\left\<\ka_1^2\right\> \sim 2 \Delta_\mathcal{R}^2 (k_\text{eq} \chi_*)^3\simeq 0.014\left(\frac{\Omega_mh^2}{0.14}\right)^3\left(\frac{\chi_*}{14\,\text{Gpc}}\right)^3,
\ee
where $\chi_*$ denotes the comoving distance to $z_*$. This is a 1\% increase for canonical cosmological parameters.
(Here and throughout we only consider the dominant contributions~-- there are smaller effects such as time delay and ISW contributions in every quantity we calculate, see discussion in Section~\ref{s:dom}. Within this approximation the relative change in the area distance and in the luminosity distance is the same, therefore we shall not systematically distinguish between them but just use the term 'distance' which can always be replaced by 'area distance'.) Naively, we can associate a perturbative change in the distance $\<d\>=d_0(1+\Delta)$ with a change in the spectrum to
\be\label{dsjbsdvb}
\tilde C_\ell=\left(\frac{d_0}{\<d\>}\right)^2 C_{\frac{d_0}{\<d\>}\ell}=(1-2\Delta)C_{\left(1-\Delta\right)\ell}\,,
\ee
where $C_\ell$ is the spectrum in the background~\cite{vonlanthen}. For $\Delta>0$ as in eq.~\eqref{ds} this gives a shift in the peaks to larger $\ell$.
Assuming that the shifted distance is measured by the CMB, this increases the estimation of $H_0$ by 5\%, while decreasing $\Omega_m$ by 10\%  \cite{cmb}. 

However, CMB observations are not directly sensitive to the expectation value of the distance but rather to its average over directions. This is closer in essence to how we extract a model from observations~-- by averaging over the sky.  The angular power spectrum of the CMB, $C_\ell$, can be approximated by averaging the temperature anisotropy over many patches of size $\pi/\ell$ in different directions of the sky. In a companion paper~\cite{dav}, we show that at second order in perturbation theory averaging over directions followed by an expectation value is not equivalent to the expectation value (followed by an average over directions). An alternative, but more relevant, mean distance is therefore
\be
\label{dsav}
\avd{d_*}\equiv \frac{1}{4\pi}\left\langle\int d\Omega_{\bn_o} d_*(\bn_o)\right\rangle=d_{0*}\Big[1-\frac{1}{2}\kap\Big]\, .
\ee  
Lensing therefore also generates a {\it decrease} in the observed \emph{angular mean} distance to the last scattering surface.  It is remarkable that although the expectation value of the distance along a single line of sight is increased, the total effect on the observers sky serves to cancel this effect, and bring the mean distance last scattering surface closer. Interpreted as in eq.~\eqref{dsjbsdvb}, this shifts the peaks to lower $\ell$.

In this paper, we show that eq.~\eqref{dsjbsdvb} is too simplistic to describe the change in the spectrum when the magnification matrix is affected by a combination of shear and convergence. We derive an approximate expression for the lensed angular power spectrum as a function of the convergence and the shear, including second-order contributions.  We show that the convergence at second order does not contribute to the averaged $C_\ell$ since it can be written as a total divergence. Hence lensing changes the averaged $C_\ell$ via the square of first-order terms only, and produces a shift to lower $\ell$.

We compare our derivation with standard CMB derivations in terms of the bending angle $\bal$. Standard expressions take into account a variety of lensing effects from terms of the form $\alpha_1^2\sim\mathcal{O}(\kappa_1^2)$, where $\bal_1$ is the first-order bending angle. We extend this calculation to the second-order bending angle $\bal_2$ and show that this term does not contribute to the averaged $C_\ell$. We conclude therefore that no second-order terms have been neglected in standard CMB analyses and that lensing cannot be responsible for the tension between local and CMB measurements of $H_0$ as suggested could be the case in~\cite{cmb}.

The remainder of the paper is organised as follow: in section~\ref{sec:Clalpha}, we present the standard derivation of the lensed $C_\ell$ in terms of the deflection angle $\bal$, extending it up to second order. In section~\ref{sec:Clkappa}, we derive an approximate expression for the lensed angular power spectrum as a function of the convergence and the shear at second order. We conclude in section~\ref{sec:conclusion}.

\section{The contribution of {\large $\bal$} to the lensed power spectrum}

\label{sec:Clalpha}

Lensing shifts the position of the points on the last-scattering surface. The lensed temperature observed in direction $\bn_o$ is therefore given by the unlensed temperature in direction $\bn$
\be
\label{T}
\tilde T(\bn_o)=T(\bn)=T(\bn_o+\bal)\simeq T(\bn_o)+\bal\cdot\bnabla T(\bn_o) +\frac{1}{2}\al^i \al^j \partial_i\partial_j T(\bn_o)+ \cdots
\ee
where  $\bal$ is the bending angle. In the current calculations of the lensed power spectra, only the first-order contribution to the deflection angle $\bal$ is taken into account, both in the second and third term on the right-hand side of eq.~\eqref{T}  \cite{Lewis:2006fu}. Here we include the effect of $\bal$ at second order $\bal=\bal_1+\bal_2/2$.

The second-order $\bal_2$ generates a new contribution to the lensed power spectrum given by the correlation of
\be
\label{deltaT2}
\delta_2 T=\bal_2\cdot\bnabla T(\bn_o), 
\ee
with the unlensed temperature. The deflection angle $\bal_2$ is derived in Appendix~\ref{app:deviation vector}. It reads
\be
\al^a_2(\chi_*, \bn_o) = 8\int_0^{\chi_*}d\chi \int_0^{\chi}d\chi'\frac{(\chi_*-\chi)(\chi-\chi')}{\chi^*}\partial_b \partial^a\Psi(\chi, \bn_o)\partial_b\Psi(\chi', \bn_o) \, ,
\ee
where $a=\theta,\varphi$ denotes the two components of $\bal$ transverse to the photon direction $\bn_o$. Here $\partial_a\equiv e_a^i\partial_i$, with $\mathbf{e}_a$ orthogonal to $\bn_o$ (see appendix C) and $\Psi$ denotes the Weyl potential defined in appendix A.

 In Appendix~\ref{a:cmb}, we show explicitly that the contributions of $\delta_2 T$ to the two-point correlation
\be
\<\delta_2 a_{\ell m}\, a_{\ell'm'} \>
\ee 
exactly vanishes if we consider $\bal_2$ and $T$ to be uncorrelated. This is a good approximation, since most of the deflection angle is generated at $z\lesssim 100$, whereas the CMB anisotropies stem from the last scattering surface (apart from the ISW term which is relevant only on large scales, dominated by cosmic variance). We derive this result both in the full-sky and in the flat-sky approximation.

Alternatively, the result can be understood by noting that
\be
\< \delta_2 T(\bn_o)T(\bn'_o)\>=\<\bal_2(\bn_o) \>\<\bnabla T(\bn_o) T(\bn'_o) \>.
\ee
Since both $\bal_2$ and $\bnabla T(\bn_o) T(\bn'_o)$ are vectors on the sphere, their expectation values have to vanish in order not to break statistical isotropy. 

This shows that if we are interested in the effect of lensing on the CMB up to second order in perturbation theory, it is sufficient to include the first-order bending angle squared. In section~\ref{sec:Clkappa} we show how this result translates in terms of the convergence and the shear. 


Going to higher order we see that $\< \delta_1T(\bn_o)\delta_2 T(\bn'_o)\>$ contains an odd number of correlators and therefore it vanishes for Gaussian initial perturbations.  The first non-vanishing higher order terms are therefore  $\< \delta_2T(\bn_o)\delta_2 T(\bn'_o)\>$ and $\< \delta_1T(\bn_o)\delta_3 T(\bn'_o)\>$,  which have been calculated in \cite{schafer}. As is to be expected, the effect on the temperature anisotropies of these terms is very small. Nevertheless, for percent accuracy in the polarization, they have to be taken into account; in particular, their effect on converting $E$- into $B$-polarization seems to be considerable. However, these results seem to contradict a lensing calculation of post-Born terms which obtains much smaller corrections~\cite{krause}; therefore, this clearly requires more work.

\section{The contributions of {\large$\kappa$} and {\large$\gamma$} to the lensed power spectrum}
\label{sec:Clkappa}

We derive now an approximate expression for the lensed angular power spectrum in terms of the shear and the convergence. We follow the approach of~\cite{bucher, bisp}, keeping all relevant terms up to second order in perturbation theory. In this approach, the convergence $\kappa$ and the shear $\gamma\equiv \go+i\gt$ are assumed to be constant over the patch of the sky in which we measure the temperature fluctuations. This approximation is well motivated to compute the effect of a large-scale lensing mode on a small-scale temperature fluctuation. It clearly breaks down when one considers fluctuations of the lensing potential at the same scales as the temperature fluctuations.  

Since we are mainly interested in small angular scales, we work in the flat-sky approximation throughout. More details on the flat-sky approximation are found in Appendix~\ref{a:flatsky}. We measure fluctuations in the temperature at two observed positions $\bx_o$ and $\by_o$ in the sky. These positions are deflected by lensing so that their true positions are $\bx$ and $\by$. 
If the separation between the points is small we have
\be
\bx-\by=A(\bx_o-\by_o) \, ,
\ee
where $A$ is the magnification matrix
\be
A=\left(\begin{array}{cc} 1-\kappa-\go& -\gt\\ -\gt& 1-\kappa+\go \end{array}\right)
= (1-\ka)\mathds{1} + \ga \, .
\ee
Perturbations in the distance are given by the determinant of $A$ via
\be
\frac{d}{d_0}=\sqrt{\det A}\simeq 1-\kappa-\frac{1}{2}|\gamma|^2\,,
\ee
where
\be
|\gamma|^2\equiv |\det\gamma|=\big(\go\big)^2 +\big(\gt\big)^2\, .
\ee

The lensed CMB power spectrum, $\tilde C(\bl)$, is then given by
\begin{align}
\label{Clens}
&\<\tilde T(\bl)\tilde T^*(\bl') \>=\int \frac{d^2\bx_o}{2\pi}\int \frac{d^2\by_o}{2\pi} e^{-i\bl\bx_o+i\bl'\by_o}\<\tilde T(\bx_o)\tilde T(\by_o)\>=
\int \frac{d^2\bx_o}{2\pi}\int \frac{d^2\by_o}{2\pi} e^{-i\bl\bx_o+i\bl'\by_o}\<T(\bx)T(\by)\>\nonumber\\
&=\int \frac{d^2\bx_o}{2\pi}\int \frac{d^2\by_o}{2\pi} e^{-i\bl\bx_o+i\bl'\by_o}\int \frac{d^2\bl_1}{2\pi}e^{i\bl_1(\bx-\by)}C(\ell_1)
=\int \frac{d^2\bx_o}{2\pi}\int \frac{d^2\by_o}{2\pi} e^{-i\bl\bx_o+i\bl'\by_o}\int \frac{d^2\bl_1}{2\pi}e^{i\bl_1A(\bx_o-\by_o)}C(\ell_1)\nonumber\\
&=\int \frac{d^2\bl_1}{2\pi} C(\ell_1)\delta(\bl-A\bl_1)\delta(\bl-\bl') =\frac{C\big(|A^{-1}\bl |\big)}{{\rm det}A}\delta(\bl-\bl')\, ,
\end{align}
where in the second equality we have used that the lensed temperature at position $\bx_o$ is given by the unlensed temperature at position $\bx$. For a change in the distance that arises from both convergence and shear then, the correct shift in the spectrum is given by
\be\label{sdkbcsjdc}
\tilde C(\ell)=\left(\frac{d_0}{d}\right)^2\,C\big(|A^{-1}\bl |\big)
\ee 
and not simply by eq.~\eqref{dsjbsdvb}, which only applies for a diagonal magnification matrix. 

Keeping terms up to second order in the convergence and the shear we have 
\bea
\label{Al}
|A^{-1}\bl |= \left[1+\kappa -\gl +\kappa^2 +\frac{3}{2}|\gamma|^2 -2\kappa\,\gl-\frac{1}{2}(\gl)^2 \right]\ell\equiv \beta \ell\, .
\eea
The lensed angular power spectrum depends therefore not only on $\ell$ but also on the direction of $\bl$. This breaking of isotropy is generated by the terms proportional to $\gl$. 

Defining $\tilde D(\ell)=\ell^2\tilde C(\ell)$, eq.~\eqref{sdkbcsjdc} becomes
\bea
\tilde D(\ell)&=&\frac{1}{\det\!A\,\beta^2} D(\beta\ell)\nonumber\\
&\simeq& \left[1+2\gl-2|\gamma|^2+4\big(\gl\big)^2+2\kappa\,\gl \right]D(\ell)
+\left[\kappa-\gl+\kappa^2+\frac{3}{2}|\gamma|^2-\frac{5}{2}\big(\gl\big)^2 \right] \ell D'(\ell)\nonumber\\
&&+\left[\frac{1}{2}\kappa^2 +\frac{1}{2}\big(\gl \big)^2-\kappa\,\gl\right] \ell^2 D''(\ell)\, .
\eea

We are interested in the isotropic part of $\tilde D(\bl)$, averaged over all directions of $\bl$. Using that
\be
\frac{1}{2\pi}\int d\theta_{\ell}\,\gl=0\quad\quad\mbox{and}\quad\quad
\frac{1}{2\pi}\int d\theta_{\ell}\big(\gl\big)^2=\frac{1}{2}|\gamma|^2\, ,
\ee
where $\theta_\ell$ denotes the direction of $\bl$, we find~\footnote{Note that the average over the direction of $\bl$ can only be performed after $D(\beta\ell)$ has been expanded. Averaging directly the argument~$\beta$ in eq.~\eqref{Al} is not consistent.}
\be
\tilde D(\ell)=D(\ell)+\left[\kappa+\kappa^2+\frac{1}{4}|\gamma|^2 \right] \ell D'(\ell)
+\frac{1}{2}\left[\kappa^2+\frac{1}{2}|\gamma|^2 \right] \ell^2 D''(\ell)\, .
\ee

So far the calculation has been done for a small patch of the sky, where $A$ was assumed to be constant. In practice the observed $\tilde D(\ell)$ are effectively averaged over the sky. The expectation value of the averaged spectrum is therefore given by 
\be
\label{Dlav}
\avd{\tilde D(\ell)}=D(\ell)+\left[\avd{\kappa}+\avd{\kappa^2}+\frac{1}{4}\avd{|\gamma|^2} \right] \ell D'(\ell)
+\frac{1}{2}\left[\avd{\kappa^2}+\frac{1}{2}\avd{|\gamma|^2} \right] \ell^2 D''(\ell)\, ,
\ee
where a bar $\overline{\phantom{x}}$ denotes an average over directions $\bn_o$, as defined in eq.~\eqref{dsav}.
 The first order convergence vanishes on average $\avd{\kappa_1}=0$. The second-order convergence $\kappa_2$ can potentially contribute to the average. In appendix~\ref{a:lens}, we calculate explicitly the convergence up to second order in perturbation theory. We show that $\kappa_2$ can be written as a total derivative 
\bea
\kappa_2=\int_0^{\chi_*}\hspace{-0.2cm} d\chi\int_0^\chi \hspace{-0.2cm} d\chi' \frac{(\chi_*-\chi)(\chi-\chi')}{\chi^2\chi'\chi_*}
&&\hspace*{-0.4cm} \Bigg[\spart\Big(\spart\spartb\Psi(\no,\chi)\spartb\Psi(\no,\chi')+\spartb^2\Psi(\no,\chi)\spart\Psi(\no,\chi')\Big)   \nonumber\\
&&+\spartb\Big(\spart\spartb\Psi(\no,\chi)\spart\Psi(\no,\chi')+\spart^2\Psi(\no,\chi)\spartb\Psi(\no,\chi') \Big) \Bigg]\, ,   \label{kappa2}
\eea
where $\Psi\equiv (\phi+\psi)/2$ and the transverse operators $\spart$ and $\spartb$ are defined in appendix~\ref{a:lens}, see also~\cite{book}. As a consequence the average of $\kappa_2$ vanishes~\footnote{Note that for second-order terms, the difference between the average over direction and the ensemble average is higher order so that we have $\avd{\kappa_2}=\<\kappa_2 \>$. Recall also that we are only considering terms with 4 transverse derivatives.}: $\avd{\kappa_2}=0$. The only contribution to the lensed spectrum comes therefore from the square of the first-order convergence and shear. In appendix~\ref{a:lens} we show also  that the combination $\kappa_1^2-|\gamma_1^2|$ can be written as a total divergence, so that on average 
\be \label{k=g}
\kap=\< |\gamma_1^2|\> \,.
\ee
With this eq.~\eqref{Dlav} becomes
\be
\label{Dlavfin}
\avd{\tilde D(\ell)}=D(\ell)+\frac{5}{4}\kap \ell D'(\ell)
+\frac{3}{4}\kap \ell^2 D''(\ell)\, .
\ee
The square of the first-order convergence affects therefore the lensed power spectrum. The first term in eq.~\eqref{Dlavfin} shifts the position of the peaks, whereas the second one both smoothes the peaks and shifts them. These two types of corrections are however already consistently included in standard CMB analyses, which include terms up to $\OO\left(\alpha_1^2\right)$, where $\bal$ is the deflection angle (see section~\ref{sec:Clalpha}). The only contribution not included in previous calculations of the lensed CMB spectrum is the second-order convergence $\kappa_2$. However, as we show here its effect exactly vanishes on average, so it does not introduce any additional change in the spectra. 

Finally let us note that if we replace eq.~\eqref{sdkbcsjdc} by expression~\eqref{dsjbsdvb}, assuming that the changes in the spectrum are due only to the mean distance $\avd{d}=d_0\big(1-\kap/2\big)$, instead of eq.~\eqref{Dlavfin} we would obtain
\bea
\label{Dlavdist}
\avd{\tilde D(\ell)}&=&  D(\ell)+\frac{1}{2}\kap \ell D'(\ell)\, .
\eea
The shift in smaller than the one given by the first term in eq.~\eqref{Dlavfin} and the smoothing term is not present.

\subsection{Shift in the peaks}

We can now calculate the shift in the position of the peaks induced by the square of the convergence. Denoting the (observed) peak of the lensed $\tilde D(\ell)$ by  $\ell_o$ and the peak of the unlensed $D(\ell)$ by  $\ell_*$, we have 
\be
\avd{\tilde D'(\ell_o)}=0=\left(1+\frac{5}{4}\kap \right) D'(\ell_o)+\frac{11}{4}\kap\ell_oD''(\ell_o)+\frac{3}{4}\kap\ell_o^2 D'''(\ell_o)\, .
\ee
Expanding the unlensed spectrum around $\ell_*$ using $\ell_o=\ell_*+\delta\ell$ we find
\be
\label{deltal}
\frac{\delta\ell}{\ell_*}\simeq-\frac{\kap}{4}\left[11+3\ell_* \frac{D'''(\ell_*)}{D''(\ell_*)} \right]\, .
\ee

The shift can be calculated by approximating the unlensed spectrum by $D(\ell)\propto \cos^2\left(\pi\ell/\ell_* \right)$.
Inserting this in eq.~\eqref{deltal} gives simply~\footnote{Note that since $D'''(\ell_*)=0$, the next term in the expansion around $\ell_*$ becomes relevant and a more accurate expression for $\delta\ell$ in this case is given by $\frac{\delta\ell}{\ell_*}\simeq-\frac{11\kap}{4}\left[1-\frac{3\kap}{4}\ell_*^2\frac{D''''(\ell_*)}{D''(\ell_*)} \right]=-\frac{11\kap}{4}\left[1+3\kap \pi^2\ell_*\right]$.}
\be
\label{deltalapp}
\frac{\delta\ell}{\ell_*}\simeq-\frac{11\kap}{4}< 0\, .
\ee
Lensing therefore shifts the position of the peaks to smaller multipoles, which is consistent with a {\it decrease} in the observed distance, as seen in eq.~\eqref{dsav}. The shift is proportional to $\kap$ which reaches a percent at the last-scattering surface. However, the present calculation of the lensed spectrum should be taken with precaution since it is valid only for large-scale lensing modes (which can be approximated as constant for a fixed $\ell$). The small-scale lensing modes contribute however significantly to $\kap$. For those modes the calculation above cannot be trusted and it is necessary to account for convolutions between the temperature and the lensing deflection. The shift calculated in eq.~\eqref{deltalapp} therefore significantly over-estimates the true shift induced by first-order terms squared, but the qualitative behaviour is sound.

Finally as mentioned above, in addition to a shift in the position of the peaks, lensing also induces a smoothing of the peaks, due to the second term in eq.~\eqref{Dlavfin}. This smoothing actually dominates over the displacement term and it constitutes the main impact of lensing on the extraction of cosmological parameters.

\subsection{Why transverse derivatives dominate}\label{s:dom}

In our derivation of the convergence and the deviation vector at second order, we have neglected all contributions with less than four (respectively three) transverse derivatives of the gravitational potential (see appendices~\ref{a:lens} and~\ref{app:deviation vector}). This is justified since the gravitational potential remains small on all scales, while its second spatial derivatives can become large, since density fluctuations are large on small scales
\be
H^{-2} \De \phi \simeq\frac{\de\rho}{\rho} \gtrsim \OO(1) \qquad \mbox{ on small scales.}
\ee
Moreover, time derivatives of the potential can be neglected with respect to spatial derivatives, since cosmological perturbations vary very slowly with time. Finally, radial derivatives on the light-cone are also smaller than transverse derivatives. We can indeed rewrite radial derivatives along the null geodesic as
\be
\int_{\la_*}^{\la_0}d\la\, \dd_r\phi  = \int_{\la_*}^{\la_0} d\la\,\bn_o\bnabla\phi  =  \int_{\la_*}^{\la_0}d\la\, (\bn-\bal)\bnabla\phi =\phi(\la_0)- \phi(\la_*) - \int_{\la_*}^{\la_0}d\la\, [\dd_t\phi+\bal\bnabla\phi ] \, , 
\ee
where $\la_*$ is the value of the affine parameter at the source. As argued before, $\phi$ and $\partial_t\phi$, are much smaller than spatial derivatives. In addition, $\bal\bnabla\phi$ is a second-order perturbation (since $\bal$ is itself a perturbation) and it is consequently smaller than transverse derivatives of $\phi$.

\section{Conclusions}

\label{sec:conclusion}

In this paper we have derived an approximate expression for the lensed angular power spectrum of the CMB in terms of the shear and the convergence. We have consistently included all (dominant) lensing terms up to second order in perturbation theory. We have shown that the pure second-order contribution proportional to $\kappa_2$ where $\ka=\ka_1+\kappa_2/2$ does not affect the lensed power spectrum, since $\kappa_2$ can be expressed as a total divergence. Also the second-order contribution to the shear only appears at third order. Squares of  first-order terms, on the other hand induce a small shift of the CMB peaks to lower multipoles. This corresponds to a decrease in the observed mean distance to the last scattering surface. We argue that this shift is properly accounted for in standard CMB analyses and that it can therefore not be responsible for the tension between local and CMB measurements of $H_0$ as suggested could be the case in~\cite{cmb}. 

The present derivation differs in two points with respect to the analysis in~\cite{cmb}. First, \cite{cmb} was using eq.~\eqref{dsjbsdvb}, which is actually modified when lensing shear is present~-- we have given the correct version in eq.~\eqref{sdkbcsjdc}, which is valid  for a non-diagonal magnification matrix. Second, \cite{cmb} computed the expectation value of the distance, whereas as we discuss in detail in~\cite{dav}, the quantity which is more relevant to the CMB power spectrum is the expectation value of the angular average of the distance. This difference is important, since lensing {\it decreases} the angular average of the distance, whereas it {\it increases} the expectation value of the distance. 


In our derivation we have included only the dominant contributions to the convergence and the deflection angle, i.e. those with the maximum number of transverse derivatives. However, since these contributions vanish on average, one can wonder if the sub-dominant ones will give an additional shift to the distance. Since the convergence is a scalar field, it can only have an even number of transverse derivatives. The terms with no transverse derivatives will certainly contribute to the mean distance (since they cannot be written with a total divergence). These terms are due for example to the Integrated Sachs-Wolfe effect or the Shapiro time-delay. They affect photon propagation and change the physical length of the geodesics between us and the last-scattering surface. Therefore it is natural that these terms change the mean distance to the CMB. However, their amplitude is about $10^{-5}$ at first order and we expect the second-order terms to be the square of this.

The terms with two transverse derivatives may or may not vanish on average. Only a full calculation of these terms can determine if they can be written with a total divergence or not.  Since these terms describe a coupling between longitudinal and transverse deflections, it is probable that they will shift the distance. However, also the impact of these terms will be very small. Eq.~\eqref{kappapp} shows that first-order terms square (with four transverse derivatives) are of the percent level, and therefore terms with only two transverse derivatives will be well below the percent, at most of the order of $10^{-6}$. (The terms with 2 transverse derivatives can schematically be written as a coupling between the first-order convergence and longitudinal perturbations, proportional to $\Psi, \dd_t{\Psi}$ or $\partial_r\Psi$ integrated along the photon trajectory. The first-order convergence reaches at most a few percent at the last-scattering surface, whereas the longitudinal perturbations are of the order of $10^{-5}$.) These subdominant contributions to the mean distance to the CMB will therefore  probably  remain undetected for a long time. 

\vspace{0.11cm}

\noindent{\it Note added:} Some similar issues are discussed in \cite{KP} which appeared after this work was completed.

\vspace{1cm}

\[\]
{\bf Acknowledgments:} It is a pleasure to thank Anthony Challinor for interesting discussions and for pointing out an error in the first version of the draft. We also thank Alex Hall, Antony Lewis,  Giovanni Marozzi and Fabian Schmidt for interesting and helpful discussions. RD is supported by the Swiss National Science Foundation. CC, RM and OU are supported by the South African National Research Foundation.
RM and OU  are supported by the South African Square Kilometre Array Project and  RM acknowledges support from the UK Science \& Technology Facilities Council (grant ST/K0090X/1). 

\appendix

\section{The lens map to second order}\label{a:lens}

We derive the lens map to second order, using the geodesic deviation equation. We follow the formalism and notation of~\cite{francis}, where the shear has been computed to second order.  Here we also compute the convergence, however, we only consider the perturbations with the maximal number of transverse derivatives. For redshifts $z\gtrsim 0.5$ these dominate the result.
We consider  scalar perturbations in longitudinal gauge, 
\be
d s^2= a^2(\eta) \left[-\big(1+2\phi\big)d \eta^2+ \big(1-2\psi \big)d \bx^2 \right]\, .
\label{metric}
\ee
Photon propagation is conformally invariant, hence we can ignore the scale factor $a(\eta)$. We also introduce the Weyl potential $\Psi=(\phi+\psi)/2$.

The geodesic deviation equation can be rewritten as an evolution equation for the $2\times 2$ magnification matrix $\DD_{ab}$
\be
\label{eqDab}
\frac{d^2}{d\lambda^2}\DD_{ab}=\mR_{ac} \DD_{cb}\, ,
\ee
where $\mR_{ac}\equiv R_{\mu\nu\rho\sigma}k^\nu k^\rho n_a^\mu n_b^\sigma$ and $\lambda$ is the affine parameter along the photon geodesic. Here $n_a^\mu$, with $a=1,2$, are two unit vectors orthogonal to $k^\mu$ and to the observer velocity $u_o^\mu$. Eq.~\eqref{eqDab} can be rewritten in terms of the conformal distance $\chi=\eta-\eta_o$. At first order, after integration by part, the solution becomes
\be
\DD_{ab} (\chi_*) = \int_0^{\chi_*} d \chi (2-k^0)\delta_{ab} + \int_0^{\chi_*} d \chi (\chi_*-\chi) \chi \mR_{ab}\, ,
\label{map_first_order}
\ee 
where $\chi_*$ is the conformal distance to the last scattering surface.
This first-order solution can be used to calculate the solution at second order. Formally, at second order we have
\be
\DD_{ab}(\chi_*) = \chi_* \delta_{ab} + \int_0^{\chi_*} d \chi \frac{\chi_* - \chi}{\chi} \mS_{ab}\,,
\label{D_*}
\ee
where the source term $\mS_{ab}$ is defined as
\be
\mS_{ab}(\chi) \equiv  \frac{\chi}{(k^0)^2} \mR_{ac} \DD_{cb} 
- \frac{\chi}{k^0} \frac{d k^0}{d \chi} (2-k^0) \delta_{ab} - \chi \frac{d k^0}{d \chi} \int_0^\chi d \chi' \chi' \mR_{ab}\,.
\label{source_inter}
\ee
At second order we have two types of terms: the second-order source terms integrated along the background trajectory, and the first-order source terms integrated along the perturbed trajectory. Expanding these first-order terms around the background trajectory
\be
\mS_{1\,ab} (x^i_{\rm pert}) = \mS_{1\,ab}(x^i) + \delta x_1^{ j} \partial_j\mS_{1\,ab}\, ,
\ee
and combining them with the second-order terms, we obtain
\be
\mS_{ab}(\chi) =  \frac{\chi}{(k^0)^2} \mR_{ac} \DD_{cb}
- \frac{\chi}{k^0} \frac{d k^0}{d \chi} (2-k^0) \delta_{ab} - \chi \frac{d k^0}{d \chi} \int_0^\chi d \chi' \chi' \mR_{ab} + \chi^2 \delta x^j \partial_j \mR_{ab} -\chi\delta x^j \partial_j\left(\frac{dk^0}{d\chi} \right) \delta_{ab} \, ,
\label{source_term}
\ee
where now all the integrals are along the background geodesic relating the image to the observer. The last two terms are the corrections to the so-called Born approximation~\footnote{Note that the last term is missing in eq.~(54) of ref.~\cite{francis}. Since it only contributes to the trace of $\DD_{ab}$ it does not affect the calculation of the shear presented there.}.

In the following, we will concentrate on the dominant contributions, i.e. those with four transverse derivatives. 
The terms in $\mR_{ab}$ have at most two transverse derivatives, those in $\delta x^i$ have at most one transverse derivative, and in $k^0$ there are no transverse derivatives. From this, we see that the only terms with four transverse derivatives in eq.~\eqref{source_term} are the first term, when both $\mR_{ac}$ and $\DD_{cb}$ are taken at first order; and the fourth term, when both $\delta x^j$ and $\mR_{ab}$ are taken at first order. At first order, the dominant contribution to $\mR_{ab}$ reads
\be
\mR_{1\,ab}=- 2e_a^i e_b^j \partial_i\partial_j\Psi\, ,
\label{R_first_order}
\ee
where the vectors $\mathbf{e}_a$, with $a=\theta,\varphi$, are orthogonal to the observed direction $\no$:
\bea
\mathbf{e}_\theta&=& (\cos \theta \cos \varphi , \cos \theta \sin \varphi, -\sin \theta)\, , \label{etheta}\\
\mathbf{e}_\varphi &=& (-\sin \varphi , \cos \varphi, 0) \, ,\label{ephi}\\
\no &=& (\sin \theta \cos \varphi , \sin \theta \sin \varphi, \cos \theta)\, . 
\eea
Inserting eq.~\eqref{R_first_order} into eq.~\eqref{map_first_order}, we obtain for the first-order dominant contribution to $\DD_{ab}$
\be
\DD_{1\,ab}(\chi) = -  2e_a^i e_b^j \int_0^{\chi} d \chi' (\chi - \chi') \chi'\partial_i\partial_j \Psi(\chi') \, .
\label{linear_mapping}
\ee
Here and in the following we drop the argument $\bn_o$ in the gravitational potential when there is no ambiguity.
With this, the first term in eq.~\eqref{source_term} becomes
\be
\mS_{2\,ab}(\chi)=4\chi e_a^ie_c^je_c^k e_b^\ell\ \dd_i\dd_j\Psi(\chi)\int_0^{\chi} d \chi' (\chi - \chi') \chi'\partial_k\partial_\ell \Psi(\chi')\, .
\ee
To calculate the contribution from the fourth term in eq.~\eqref{source_term}, we need the deviation vector $\delta x^i$ at first order. The calculation of the deviation vector is presented in appendix~\ref{app:deviation vector}. Combining eq.~\eqref{dxa} with eq.~\eqref{R_first_order}, we find for the fourth term in eq.~\eqref{source_term}
\be
\mS_{2\,ab}(\chi)=4\chi^2e_c^i\dd_i\left(e_a^je_b^\ell \dd_j\dd_\ell\Psi(\chi) \right) \cdot \int_0^{\chi} d \chi' (\chi - \chi')e_c^k\dd_k\Psi(\chi')\, .
\ee
The dominant contributions to $\DD_{ab}$ at second order then read
\begin{align}
\label{D2}
\DD_{2\,ab}(\chi_*)=4\int_0^{\chi_*}d\chi\int_0^\chi d\chi' (\chi_*-\chi)(\chi-\chi'&)
\Big[\chi'e_a^i e_c^j \dd_i\dd_j\Psi(\chi)e_c^k e_b^\ell \dd_k\dd_\ell \Psi(\chi') \\
& +\chi e_a^i e_b^j e_c^k \dd_i\dd_j\dd_k\Psi(\chi)e_c^\ell \dd_\ell \Psi(\chi')\Big]\, ,\nonumber
\end{align}
where we have neglected derivatives of the vector $e_a^i$, which lead to contributions with less transverse derivatives.

The transverse derivatives $e_a^i\dd_i$ can be rewritten in terms of the derivatives on the sphere $\spart$ and $\spartb$. These operators depend explicitly on the spin of the field ${}_s X$ to which they are applied (for more details see e.g. appendix B of~\cite{francis}). In terms of the angles $\theta$ and $\varphi$, they read
\be
\spart \; {}_s X   \equiv - \sin^s \theta (\partial_\theta + i \csc \theta \partial_\varphi) 
(\sin^{-s} \theta)
\; {}_s X \;, 
\qquad \spartb \; {}_s X   \equiv - \sin^{-s} \theta (\partial_\theta - i \csc \theta \partial_\varphi) (\sin^{s} \theta) \; {}_s X \;.
\ee
Defining the vectors $\mathbf{e}_{\pm}=\mathbf{e}_\theta\pm i\mathbf{e}_\varphi$, we have the following relations for a scalar field $X=X(\chi)$
\be
e_+^i \dd_i X =- \frac{1}{\chi} \spart X \,, \qquad 
e_-^i \dd_i X = - \frac{1}{\chi} \spartb X\,.
\label{e+}
\ee
Then, using that $\chi e_\pm^i \dd_i e_\pm^j = \cot \theta e_\pm^j$ we find
\be
e_+^i e_+^j \dd_i\dd_j X = \frac{1}{\chi^2} \spart^2 X   \,, \qquad
e_-^i e_-^j \dd_i \dd_j X = \frac{1}{\chi^2} \spartb^2 X  \, ,
\label{e+e+}
\ee
and, analogously with $\chi e_\mp^i \partial_i e_\pm^j = - \cot \theta e_\pm^j - 2 n_o^j$, we have
\be
e_+^{\ i}e_-^{\ j} \dd_i\dd_j X = 
e_-^{\ i} e_+^{\ j} \dd_i\dd_j X = \frac{1}{\chi^2}  \spartb \spart  X - \frac{2}\chi \dd_r X   \, .
\label{e+e-}
\ee
Since we are only interested in the contributions with the most transverse derivatives, the last term in eq.~\eqref{e+e-} can be neglected.

We then decompose the magnification matrix in terms of the shear, the convergence and the rotation
\be
\label{Dab}
\DD_{ab}=\lambda_*\left(\begin{array}{cc}1-\kappa-\go&-\gt-\omega\\ -\gt+\omega& 1-\kappa+\go \end{array} \right)\, ,
\ee
where each component has a first and second-order part $\kappa=\kappa_1+\kappa_2/2$ and similarly for $\gamma$ and $\omega$.
At first order, from eq.~\eqref{linear_mapping} and using that $e_a^i e_a^j=(e_+^ie_-^j+e_-^ie_+^j)/2$, we obtain
\bea
\kappa_1(\chi_*)&=&\int_0^{\chi_*}d\chi\frac{\chi_*-\chi}{\chi\chi_*}\spart\spartb\Psi(\chi)\, ,\label{kappa1}\\
\gamma_1(\chi_*)\equiv \go_1+i\gt_1&=&\int_0^{\chi_*}d\chi\frac{\chi_*-\chi}{\chi\chi_*}\spart^2\Psi(\chi)\, . \label{gamma1}
\eea
The rotation $\omega$ exactly vanishes at first order.

At second order, using eq.~\eqref{D2}, we obtain 
\bea
\label{kappa2app}
\kappa_2(\chi_*)=\int_0^{\chi_*}d\chi\int_0^\chi d\chi' \frac{(\chi_*-\chi)(\chi-\chi')}{\chi^2\chi'\chi_*}
&&\Bigg[ \spart^2\Psi(\chi)\spartb^2\Psi(\chi')+\spartb^2\Psi(\chi)\spart^2\Psi(\chi')
+2\spart\spartb\Psi(\chi)\spart\spartb\Psi(\chi')\nonumber\\
&&+2\spartb^2\spart\Psi(\chi)\spart\Psi(\chi')
+2\spart^2\spartb\Psi(\chi)\spartb\Psi(\chi')  \Bigg]\, .
\eea
These terms can be combined and rewritten with a total transverse derivative, as shown in eq.~\eqref{kappa2}.

Moreover the combination $\kappa^2-|\gamma|^2$ can be computed from eqs.~\eqref{kappa1} and \eqref{gamma1}. It can also be written in terms of total transverse derivative
\bea
\label{kappasquare}
\big(\kappa_1^2-|\gamma_1|^2\big)(\chi_*)
&=&\frac{1}{2}\int_0^{\chi_*}\hspace{-0.2cm} d\chi\int_0^{\chi_*}\hspace{-0.2cm} d\chi' \frac{(\chi_*-\chi)(\chi_*-\chi')}{\chi\chi'\chi^2_*}
\Bigg[\spart\Big( \spartb\Psi(\chi)\spart\spartb\Psi(\chi')-\spart\Psi(\chi)\spartb^2\Psi(\chi')\Big)\nonumber\\
&&\hspace{5.3cm}+ \spartb\Big( \spart\Psi(\chi)\spart\spartb\Psi(\chi')-\spartb\Psi(\chi)\spart^2\Psi(\chi')\Big)\Bigg]\, . \label{kappasquare}
\eea

\section{Explicit calculation of $\kap$}\label{a:kappa1sq}

The average of the first order convergence squared can be written as
\be
\label{e:kappa}
\big\langle\kappa_1^2 \big\rangle=
\int_0^{\chi_*}\!\!\!d\chi\int_0^{\chi_*}\!\!\!d\chi'\frac{(\chi_*-\chi)(\chi_*-\chi')}{\chi_*^2\chi'\chi}
\left\langle \spart\spartb\Psi (\no,\chi)\spart\spartb\Psi(\no,\chi')\right\rangle\, .
\ee
We can expand the gravitational potential in spherical harmonics
\be
\label{psi}
\Psi(\no,\chi)=\frac{1}{2\pi^2}\sum_{\ell m}i^\ell \int d^3k T_\Psi(k,\chi)\zeta(\bk)j_\ell(k\chi)Y_{\ell m}(\no)Y^*_{\ell m}(\hbk)\, ,
\ee
where $\zeta(\bk)$ is the primordial curvature perturbation and $T_\Psi(k,\chi)$ is the transfer function for the gravitational potential. Using that 
\be
\langle \zeta(\bk)\zeta^*(\bk')\rangle=(2\pi)^3\frac{A}{k^3}\left(\frac{k}{k_\lambda}\right)^{n_s-1}\delta(\bk-\bk')\, ,
\ee
where $A$ is the primordial amplitude, $n_s$ the spectra index and $k_\lambda$ the pivot scale, and applying the derivative operators on the spherical harmonics
\be
\spart\spartb Y_{\ell m}(\no)=-\ell(\ell+1)Y_{\ell m}(\no)\, ,
\ee
we obtain 
\bea
\label{dACMB}
\langle\ka_1^2\rangle&=&\frac{A}{2\pi^2}\sum_{\ell}(2\ell+1)\ell^2(\ell+1)^2\int \frac{dk}{k}\left(\frac{k}{k_\lambda}\right)^{n_s-1}\left[\int_0^{\chi_*}d\chi\frac{\chi_*-\chi}{\chi\chi_*}T_\Psi(k,\chi)j_\ell(k\chi)\right]^2\, .
\eea

\section{Deviation vector up to second order}

\label{app:deviation vector}

The deviation vector $\delta x^\alpha$ is by construction the difference between the position on the perturbed photon geodesic and the position on the background geodesic, relating the image to the observer
\be
\delta x^\alpha\equiv x^\alpha- x_0^\alpha=\delta x^\alpha_1+\frac{1}{2}\delta x^\alpha_2\, .
\ee
The position $x^\alpha$ can be calculated from the null geodesic equation
\be
\label{geo}
\frac{d^2 x^\alpha}{d\lambda^2}+\Gamma^\alpha_{\mu\nu}k^\mu k^\nu=0\, .
\ee
Rewriting $\lambda$ in terms of the conformal distance $\chi=\eta_0-\eta$ and using that
\be
\frac{d}{d\lambda}=-k^0\frac{d}{d\chi}\, .
\ee
we obtain 
\be
\label{kevol}
\frac{d^2 x^\alpha}{d\chi^2}=-\frac{dk^0}{d\chi}\frac{k^\alpha}{(k^0)^2}-\frac{\Gamma^\alpha_{\mu\nu}k^\mu k^\nu}{(k^0)^2}\, .
\ee

\subsection{Order zero}

At zeroth order, $\Gamma^\alpha_{\mu\nu}=0$, we have
\be
\frac{dk_0^\alpha}{d\lambda}=\frac{d k_0^\alpha}{d\chi}=0 \quad\Rightarrow\quad  k_0^\alpha=\rm{cst}
\ee
We choose $k_0^0=1$ and $ k_0^i=n_o^i$. The position on the background geodesic is then
\be
 x_0^i(\chi_*)=\chi_*\cdot n_o^i\, ,\qquad
x_0^0(\chi_*)= \chi_*\, ,
\ee
where we have placed the origin at the observer $x^\alpha(0)= x_0^\alpha(0)=0$.

\subsection{First order}

At first order, \eqref{kevol} reads
\be
\frac{dk^\alpha}{d\chi}=-\frac{\Gamma^\alpha_{\mu\nu}k^\mu k^\nu}{k^0}=-\Gamma^{\alpha}_{1\,\mu\nu} k_0^\mu  k_0^\nu\, .
\ee
Integrating we obtain
\bea
k^i(\chi_*)&=&n_o^i \big(1+2\psi(\chi_*)\big)-2\int_0^{\chi_*}d\chi\partial^i\Psi(\chi) \label{ki}\, ,\\
k^0(\chi_*)&=&1-2\phi(\chi_*)+2\int_0^{\chi_*}d\chi \dd_\chi \Psi(\chi)\label{k0}\, ,
\eea
where we have neglected the perturbations at the observer. Here $\dd_\chi$ denotes a partial derivative with respect to $\chi$.

We then integrate one more time to find the position $x^\alpha(\chi_*)$:
\be
\frac{dx^\alpha}{d\chi}=\frac{k^\alpha}{k^0}
\ee
gives
\bea
x^i(\chi_*)&=&x_0^i(\chi_*)+4n_o^i \int_0^{\chi_*}d\chi\Psi-2\int_0^{\chi_*}d\chi(\chi_*-\chi)
\Big[\partial^i\Psi+n_o^i \dd_\chi \Psi \Big]\, , \label{xfirst}\\
x^0(\chi_*)&=&\chi_*= x_0^0(\chi_*)\, .
\eea
We introduce the two transverse vectors $\mathbf{e}_a=\mathbf{e}_\theta$ or $\mathbf{e}_\varphi$ which are orthogonal to the observed direction $\no$,
\bea
\mathbf{e}_\theta&=& (\cos \theta \cos \varphi , \cos \theta \sin \varphi, -\sin \theta)\, , \label{etheta}\\
\mathbf{e}_\varphi &=& (-\sin \varphi , \cos \varphi, 0) \, ,\label{ephi}\\
\no &=& (\sin \theta \cos \varphi , \sin \theta \sin \varphi, \cos \theta)\, . 
\eea
With this the deflection vector $\delta x^i$ can be split into its radial part $\delta x^r=n_{o\,i} \delta x^i$ and its transverse part $\delta x^a =e^a_i \delta x^i$. From \eqref{xfirst}, we see that the transverse part contains one spatial gradients and consequently dominates over the radial part which has no gradient since the radial part of last term in \eqref{xfirst} is a total derivative. For the transverse part
and the deflection angle we obtain to first order
\bea
\label{dxa1app}
\delta x_1^a &=&-2\int_0^{\chi_*}d\chi (\chi_*-\chi) e^{a\,i} \dd_i\Psi(\chi) \\
\label{dxa1app}
\al^a_1 &=& -2\int_0^{\chi_*}d\chi \frac{\chi_*-\chi}{\chi_*} e^{a\,i} \dd_i\Psi(\chi)
\, .
\eea

\subsection{Second order}

At second order the geodesic equation is
\be
\frac{d^2 x^i}{d\chi^2}=-\frac{dk^0}{d\chi}\frac{k^i}{(k^0)^2}-\frac{\Gamma^i_{\mu\nu}k^\mu k^\nu}{(k^0)^2}\, .
\ee
The first term on the right-hand side is negligible since it has less transverse derivatives than the second term. Indeed, at first order $dk^0/d\chi$ has no spatial derivative (see \eqref{k0}), and $k^i$ as only one transverse derivative (see \eqref{ki}). In addition, when $dk^0/d\chi$ is taken at second order, it is multiplied by $n_o^i$ and therefore does not contribute to the transverse deviation $\delta x^a$.

Neglecting the first term, we obtain
\be
x^i(\chi_*)= x_0^i(\chi_*)-\int_0^{\chi_*}d\chi\int_0^\chi d\chi' \frac{\Gamma^i_{\mu\nu}k^\mu k^\nu}{(k^0)^2}
= x_0^i(\chi_*)+\int_0^{\chi_*}d\chi(\chi-\chi_*) \frac{\Gamma^i_{\mu\nu}k^\mu k^\nu}{(k^0)^2}\, ,\label{xi2}
\ee
where we neglect the perturbations at the observer. In \eqref{xi2}, the integral is still performed on the perturbed geodesic. We rewrite this integral on the background geodesic by Taylor expanding the integrand around the background position. We get
\be
\delta x_2^{i}(\chi_*)=2\int_0^{\chi_*}d\chi(\chi-\chi_*)\left[\left( \frac{\Gamma^i_{\mu\nu}k^\mu k^\nu}{(k^0)^2}\right)^{(2)}
+\delta x_1^{\, j}\partial_j \left(\frac{\Gamma^i_{\mu\nu}k^\mu k^\nu}{(k^0)^2}\right)^{(1)}\right]\, . \label{xi2born}
\ee

Let us look at the first term
\be
\left( \frac{\Gamma^i_{\mu\nu}k^\mu k^\nu}{(k^0)^2}\right)^{(2)}=\Gamma^{i}_{2\,\mu\nu}k_0^\mu  k_0^\nu
+\Gamma^{i}_{1\,\mu\nu}\left( \frac{k^\mu k^\nu}{(k^0)^2}\right)^{(1)}\, .\label{chris}
\ee
The second-order Christoffel symbols contain terms with one spatial derivative of the second-order gravitational potential. These terms are taken into account in standard lensing analyses, by using the halo-fit power spectrum in the first order expression for $\delta x^a$. In addition, the second-order Christoffel symbols contain coupling terms of the form $\psi\partial_i \psi$, with at most one spatial derivative. We can therefore neglect them. The second term in \eqref{chris} has at most two transverse derivatives, one in the first-order Christoffel symbols and one in the first order $k_1^{i}$. It is therefore also negligible. The only relevant contribution comes therefore from the correction to the Born approximation, i.e. the second term in \eqref{xi2born}. We have
\be
\left(\frac{\Gamma^i_{\mu\nu}k^\mu k^\nu}{(k^0)^2}\right)^{(1)}=\Gamma^{i}_{1\,\mu\nu} k_0^\mu k_0^\nu=
\partial^i(\phi+\psi)-2 k_0^i \frac{d\psi}{d\chi}\, .
\ee
Neglecting the second term we obtain for the deflection
\be
\delta x_2^i(\chi_*)=-4\int_0^{\chi_*} d\chi (\chi_*-\chi)\delta x_1^j\partial_j\partial^i\Psi(\chi)\, .
\ee
The transverse part of $\delta x_2^i(\chi_*)$ is given by
\be
\delta x_2^a(\chi_*)=e^a_i\delta x_2^i(\chi_*)=8e^a_i \int_0^{\chi_*}d\chi(\chi_*-\chi)\partial_j\partial^i\Psi(\chi)
\int_0^{\chi}d\chi' (\chi-\chi')\partial^j\Psi(\chi')\, .\label{xa}
\ee
The derivative $\partial^j$ can be expanded on the basis $(\no, \bfe_\theta, \bfe_\varphi)$
\be
\partial^j=n_o^j n_o^k\partial_k + e_a^j e_a^k\partial_k\, .
\ee
With this
\be
\partial^j\Psi(\chi)\partial_j\Psi(\chi')=n_o^k\partial_k\Psi(\chi)n_o^\ell\partial_\ell\Psi(\chi')
+e_b^k\partial_k\Psi(\chi)e^{b\, \ell}\partial_\ell\Psi(\chi')\, .
\ee
Neglecting the first term, which has only radial derivatives, the transverse part of the second-order deviation vector, $\de\bx_2$, and the second-order deflection angle, $\bal_2$, become 
\bea
\delta x^a_2(\chi_*) &=&8\int_0^{\chi_*}d\chi \int_0^{\chi}d\chi'(\chi_*-\chi)(\chi-\chi')e^a_ie_b^ke^{b\,j}[\partial_k \partial^i\Psi(\chi)]\partial_j\Psi(\chi') \label{dxa}\\
\al^a_2(\chi_*) &=& 8\int_0^{\chi_*}d\chi \int_0^{\chi}d\chi'\frac{(\chi_*-\chi)(\chi-\chi')}{\chi^*}e^a_ie_b^ke^{b\,j} [\partial_k \partial^i\Psi(\chi)]\partial_j\Psi(\chi') \, .
\eea
The radial part $\delta x_2^r=n_{o\, i}\delta x_2^i$ has, by construction, less transverse derivatives and is therefore negligible with respect to the transverse part $\delta x_2^a$.

\section{Details of the calculation of the lensed CMB $\tilde C_\ell$}\label{a:cmb}

\subsection{$\tilde a_{\ell m}$ in the full-sky}

We now use the deviation vector to calculate the full-sky $\tilde a_{\ell m}$ at second order. The contribution from \eqref{deltaT2} reads
\be
\label{alm2}
\delta_2 a_{\ell m}(\no)=8\int d\Omega_{\no} Y_{\ell m}(\no)\int_0^{\chi_*}d\chi\int_0^{\chi}d\chi'(\chi_*-\chi)(\chi-\chi')
e^{ai}e_b^j\partial_i\partial_j\Psi(\no,\chi)e^{bk}\partial_k\Psi(\no,\chi')e_a^m\partial_m T(\no,\chi_*)\, .
\ee
Using that
\be
e^{ai}e_a^me_b^je^{bk}=\frac{1}{4}(e_+^ie_-^m+e_+^me_-^i)(e_+^je_-^k+e_+^je_-^k)\, ,
\ee
and the relations~\eqref{e+}, ~\eqref{e+e+} and~\eqref{e+e-} we obtain
\begin{align}
\label{alm2spart}
\delta_2& a_{\ell m}(\no)=2\int d\Omega_{\no} Y_{\ell m}(\no)\int_0^{\chi_*}d\chi\int_0^{\chi}d\chi'\frac{(\chi_*-\chi)(\chi-\chi')}{\chi^2\chi'\chi_*}\Big[\spart^2\Psi(\no,\chi)\spartb\Psi(\no,\chi')\spartb T(\no,\chi_*)\\
&+\spart\spartb\Psi(\no,\chi)\spart\Psi(\no,\chi')\spartb T(\no,\chi_*)
+\spart\spartb\Psi(\no,\chi)\spartb\Psi(\no,\chi')\spart T(\no,\chi_*)+\spartb^2\Psi(\no,\chi)\spart\Psi(\no,\chi')\spart T(\no,\chi_*) \Big]\, .\nonumber
\end{align}
To calculate the angular power spectrum, we need to cross-correlate $\delta_2 a_{\ell m}$ with the unlensed $a_{\ell'm'}$
\be
a_{\ell' m'}(\no')=\int d\Omega_{\no'} Y_{\ell' m'}(\no')T(\no',\chi_*)\, .
\ee
Looking first at the first term in the square bracket of \eqref{alm2spart} we obtain
\bea
\langle\delta_2 a_{\ell m}(\no) a_{\ell' m'}(\no')\rangle&=&2\int d\Omega_{\no} Y_{\ell m}(\no)\int d\Omega_{\no'} Y_{\ell' m'}(\no')
d\chi\int_0^{\chi}d\chi'\frac{(\chi_*-\chi)(\chi-\chi')}{\chi^2\chi'\chi_*}\\
&&\hspace{1.5cm}\times\left\langle\spart^2\Psi(\no,\chi)\spartb\Psi(\no,\chi') \right\rangle \left\langle \spartb T(\no,\chi_*)T(\no',\chi_*) \right\rangle\, ,\nonumber
\eea
where we have neglected the cross-correlation between the temperature and the gravitational potential coming from the integrated Sachs-Wolfe. The two-point function of the potential can be calculated using \eqref{psi} and the properties of the operators $\spart$ and $\spartb$ on the $Y_{\ell m}(\no)$. We get
\bea
\left\langle\spart^2\Psi(\no,\chi)\spartb\Psi(\no,\chi') \right\rangle&=&-\frac{2A}{\pi}\sum_\ell \sqrt{\frac{(\ell+2)!(\ell+1)!}{(\ell-2)!(\ell-1)!}}
\int\frac{dk}{k}\left(\frac{k}{k_\lambda} \right)^{n_s-1}T_\Psi(k,\chi)T_\Psi(k,\chi')j_\ell(k\chi)j_\ell(k\chi')\\
&&\times\sum_m (-1)^m {}_2Y_{\ell m}(\no){}_{-1}Y_{\ell m}(\no)\, . \nonumber
\eea
The sum over $m$ can be simplified using
\be
\sum_m (-1)^m {}_2Y_{\ell m}(\no){}_{-1}Y_{\ell m}(\no)=-\sum_m {}_2Y_{\ell m}(\no){}_{1}Y^*_{\ell m}(\no)=-\,{}_2Y_{\ell( -1)}(0,0)=0\quad \forall \ell\, .
\ee
The same kind of calculation applies to the other terms in \eqref{alm2spart} leading to
\be
\langle\delta_2 a_{\ell m}(\no) a_{\ell' m'}(\no')\rangle=0\, .
\ee

\subsection{$\tilde C_\ell$ in the flat-sky approximation}
\label{a:flatsky}
Let us also briefly present the calculation of the lensed $\tilde C_\ell$ in the flat-sky approximation, where the formalism is much simpler.  As we will see, in the flat-sky approximation we find that the diagonal part of the two-point function vanishes in average. The off-diagonal part on the other hand vanishes under Limber approximation. But since in the flat-sky approximation the Limber approximation becomes exact, this shows that the flat-sky result is consistent with the full-sky result above. 

In the flat-sky approximation, the $\tilde a_{\ell m}$ are replaced by 2D multipoles
\be
\tilde T(\bl)=\int \frac{d^2\xo}{2\pi}\tilde T(\xo)e^{-i\bl\cdot\xo}\, ,
\ee
where $\xo$ is the 2-dimensional observed position on the last scattering surface and $\bl$ is the 2D variable of its Fourier transform. We want to calculate the contribution from the transverse part of $\delta_2 T$ in \eqref{deltaT2} (remember that the radial part is negligible)
\be
\delta_2 T(\bl)=\int \frac{d^2\xo}{2\pi}\delta x^a\partial_aT(\xo)e^{-i\bl\cdot\xo}\, .
\ee
Using \eqref{dxa} for the second-order deflection vector and Fourier transforming (in 2D) the gravitational potential and the temperature, we obtain
\bea
\delta_2 T(\bl)&=& 8\int  \frac{d^2\xo}{2\pi}e^{-i\bl\cdot\xo}\int_0^{\chi_*}d\chi \int_0^\chi d\chi'
\frac{(\chi_*-\chi)(\chi-\chi')}{\chi_*\chi\chi'^2}\int\frac{d^2\bl_1}{2\pi}\int\frac{d^2\bl_2}{2\pi}\int\frac{d^2\bl_3}{2\pi}\nonumber\\
&&e^{i\left(\bl_1\frac{\chi}{\chi_*}+\bl_2\cdot \frac{\chi'}{\chi_*}+\bl_3\right)\cdot \xo}(\bl_1\cdot\bl_2)(\bl_1\cdot\bl_3)\Psi(\bl_1,\chi)\Psi(\bl_2,\chi')T(\bl_3,\chi_*)\, ,
\label{T2}
\eea
where we have used that in the flat-sky approximation the positions at which the potential is evaluated, $\bx$ and $\bx'$, are related to the observed position $\xo$ through
\be
\bx=\frac{\chi}{\chi_*}\xo \quad \mbox{and} \quad \bx'=\frac{\chi'}{\chi_*}\xo\, .
\ee
The integral over $\xo$ then gives
\be
\int \frac{d^2\xo}{2\pi}e^{i\left(\bl_3-\bl+\bl_1\frac{\chi}{\chi_*}+\bl_2\frac{\chi'}{\chi_*}\right)\cdot\xo}
=2\pi\delta\left(\bl_3-\bl+\bl_1\frac{\chi}{\chi_*}+\bl_2\frac{\chi'}{\chi_*} \right)\, .
\ee
With this \eqref{T2} becomes
\bea
\delta_2 T(\bl)&=&8\int_0^{\chi_*}d\chi \int_0^\chi d\chi'\frac{(\chi_*-\chi)(\chi-\chi')}{\chi_*\chi\chi'^2}\left(\frac{\chi_*}{\chi} \right)^3\int\frac{d^2\bl_2}{2\pi}\int\frac{d^2\bl_3}{2\pi}
\left[\left(\bl-\bl_3-\bl_2\frac{\chi'}{\chi_*} \right)\cdot\bl_2\right]\,\left[\left(\bl-\bl_3-\bl_2\frac{\chi'}{\chi_*} \right)\cdot\bl_3\right]\nonumber\\
&&\Psi\left(\Big(\bl-\bl_3-\bl_2\frac{\chi'}{\chi_*} \Big)\frac{\chi_*}{\chi},\chi \right)\Psi(\bl_2,\chi')T(\bl_3,\chi_*)\, .
\eea

We then cross-correlate $\delta_2 T(\bl)$ with the unlensed multipole $T(\bl')$
\bea
\langle \delta_2 T(\bl)T^*(\bl')\rangle&=&
8 \int_0^{\chi_*}d\chi \int_0^\chi d\chi'\frac{(\chi_*-\chi)(\chi-\chi')}{\chi_*\chi\chi'^2}\left(\frac{\chi_*}{\chi} \right)^3
\int\frac{d^2\bl_2}{2\pi}\int\frac{d^2\bl_3}{2\pi}\left[\left(\bl-\bl_3-\bl_2\frac{\chi'}{\chi_*} \right)\cdot\bl_2\right]\\
&&\times \left[\left(\bl-\bl_3-\bl_2\frac{\chi'}{\chi_*} \right)\cdot\bl_3\right]\,
\langle T(\bl_3)T^*(\bl')\rangle\left\langle\Psi\left(\Big(\bl-\bl_3-\bl_2\frac{\chi'}{\chi_*} \Big)\frac{\chi_*}{\chi},\chi \right)\Psi(\bl_2,\chi')\right\rangle\, .\nonumber
\eea
Using that
\be
\langle T(\bl_3)T^*(\bl')\rangle=C^{T}(\ell')\delta(\bl_3-\bl')\, \qquad \ell' =|\bl'|\,,
\ee
and
\be
\left\langle\Psi\left(\Big(\bl-\bl_3-\bl_2\frac{\chi'}{\chi_*} \Big)\frac{\chi_*}{\chi},\chi \right)\Psi(\bl_2,\chi')\right\rangle
=C^\Psi(\ell_2, \chi, \chi')\delta\left(\bl_2+\left(\bl-\bl_3-\bl_2\frac{\chi'}{\chi_*} \right)\frac{\chi_*}{\chi} \right)\, \qquad \ell_2 =|\bl_2|\,,,
\ee
we obtain for the two-point correlation function
\bea
\langle \delta_2 T(\bl)T^*(\bl')\rangle&=&
8 C^{T}(\ell')\int_0^{\chi_*}d\chi \int_0^\chi d\chi'\frac{(\chi_*-\chi)(\chi-\chi')}{\chi_*\chi\chi'^2}\left(\frac{\chi_*}{\chi} \right)^3
\int\frac{d^2\bl_2}{(2\pi)^2}\left(\bl-\bl'-\bl_2\frac{\chi'}{\chi_*} \right)\cdot\bl_2\,\left(\bl-\bl'-\bl_2\frac{\chi'}{\chi_*} \right)\cdot\bl'\nonumber\\
&& C^\Psi(\ell_2, \chi,\chi')\delta\left(\bl_2\frac{\chi-\chi'}{\chi}+(\bl-\bl')\frac{\chi_*}{\chi} \right) \, .
\label{Cfin}
\eea
The diagonal part of the two-point function vanishes. Indeed when $\bl=\bl'$, the Dirac-delta function forces either $\chi=\chi'$, in which case \eqref{Cfin} vanishes due to the kernel $\chi-\chi'$ or $\bl_2=0$, in which case \eqref{Cfin} also vanishes, due to the factor $\left(\bl-\bl'-\bl_2\frac{\chi'}{\chi_*} \right)\cdot\bl_2$. 

The non-diagonal part $\bl\neq \bl'$ vanishes due to Limber approximation. 
The angular power spectrum of the gravitational potential is indeed given by
\be
C^\Psi(\ell_2, \chi,\chi')=4\pi \int \frac{dk}{k} P_{\rm in}(k) T_\Psi(k,\chi)T_\Psi(k,\chi')j_{\ell_2}(k\chi)j_{\ell_2}(k\chi')\, ,\label{Cpsi}
\ee
where $P_{\rm in}$ is the initial power spectrum and $T_\Psi$ is the transfer function.
In Limber approximation the integral over $k$ is simplified using
\be
\int dk k^2 f(k)j_l(k\chi)j_\ell(k\chi')\simeq \frac{\pi}{2\chi^2}\delta(\chi-\chi')f\left(\frac{\ell}{\chi}\right)\, ,
\ee
and the non-diagonal contribution vanishes due to the factor $\chi-\chi'$. The flat-sky result is therefore consistent with the full-sky result.


\end{document}